\documentstyle[epsfig]{aa}

\begin{document}

\title{Strong GeV Emission Accompanying TeV Blazar H1426+428}
\author{Y. Z. Fan\inst{1,2,3}, Z. G. Dai\inst{1} and D. M.
Wei\inst{2,3}}

\offprints {Y. Z.~Fan}

\mail{yzfan@pmo.ac.cn}

\institute{Department of Astronomy, Nanjing
University, Nanjing, 210093, China \and Purple Mountain
Observatory, Chinese Academy of Sciences, Nanjing, 210008, China
\and National Astronomical Observatories, Chinese Academy of
Sciences, Beijing, 100012, China}
\date{Received date September 8, 2003/ Accepted date October 30, 2003}
\titlerunning{Strong GeV Emission accompanying
TeV Blazar H1426+428}

\abstract{For High frequency BL Lac objects (HBLs) like H1426+428,
a significant fraction of their TeV gamma-rays emitted are likely
to be absorbed in interactions with the diffuse IR background,
yielding $e^\pm$ pairs. The resulting $e^\pm$ pairs generate one
hitherto undiscovered GeV emission by inverse Compton scattering
with the cosmic microwave background photons (CMBPs). We study
such emission by taking the 1998-2000 CAT data, the reanalyzed
1999 \& 2000 HEGRA data and the corresponding intrinsic spectra
proposed by Aharonian et al. (2003a). We numerically calculate the
scattered photon spectra for different intergalactic magnetic
field (IGMF) strengths. If the IGMF is about $10^{-18}{\rm G}$ or
weaker, there comes very strong GeV emission, whose flux is far
above the detection sensitivity of the upcoming satellite GLAST!
Considered its relatively high redshift ($z=0.129$), the detected
GeV emission in turn provides us a valuable chance to calibrate
the poor known spectral energy distribution of the intergalactic
infrared background, or provides us some reliable constraints on
the poorly known IGMF strength.

}
\maketitle

\section{Introduction}
Several High frequency BL Lac objects (HBLs) such as Mkn501,
Mkn421, PKS 2155-304, 1ES 2344$+$514, H1426$+$428 and
1ES1959$+$650 are of great interest, since they emit TeV photons
(Catanese \& Weekes 1999;Aharonian et al. 1999, 2002a; Horns
2002). As early as in 1960s, it was pointed out that the observed
high energy gamma photon spectra from TeV sources might be
significantly modified---as the original high energy gamma-rays
satisfying $E_{\rm \gamma}E_{\rm B}>2(m_{\rm e}c^2)^2$ (where
$E_{\gamma}$ is the energy of the seed high energy gamma-ray,
$E_{\rm B}$ is the infrared$-$UV background photon) travel through
the intergalactic infrared$-$UV background, a significant fraction
of them will be absorbed, leading to $e^\pm$ pairs (Nikishov 1962;
Gould \& Schr\'{e}der 1965; see also Stecker, de Jager \& Salamon
1992). If the optical depth to TeV photons is evaluated, the
intrinsic spectrum can be derived. Such calculations were firstly
made for Mkn501 (during the 1997 flaring activity), and then for
H1426+428, indicating that their intrinsic hight-energy spectra
have broad, flat peaks which are much higher than the observed one
in the $\sim 5-10$ TeV range (e.g. Konopelko et al. 1999;
Aharonian, Timokhin \& Plyasheshnikov 2002; de Jager \& Stecker
2002; Aharonian et al. 2003a). Very recently, Dai et al. (2002)
have suggested that inverse Compton (IC) scattering of the
resulting $e^\pm$ pairs against the cosmic microwave background
photons (CMBPs) may produce a new GeV emission component in TeV
blazars \footnote{For Gamma-ray bursts, such Compton scattering
leads to an observable, delayed MeV-GeV emission component. In
fact, there are more than six events have been detected, the well
known one is the GRB 940217, whose delayed MeV$-$GeV emission
lasts for $\sim$ 5000{\rm s} (Hurley et al. 1994). If the model
proposed by Plaga (1995) and Cheng \& Cheng (1996) (see also Dai
\& Lu 2002; Wang et al. 2003) is correct, the intergalactic
magnetic field is $\approx 10^{-20}{\rm G}$ (For GRB 940217, the
redshift is unavailable. But with the empirical relation
established based on several GRBs with known redshifts, a rough
estimate suggests its redshift is $\sim0.68$ (Fenimore \&
Rumirez-Ruiz 2000). On the other hand, the typical energy of the
observed delayed photons is $50{\rm MeV}$, which implies the
typical energy for the high energy seed photons $E_{\rm
\gamma}\approx 0.3~{\rm TeV}$. Substituting these two estimates
into equation (3), we have $B_{\rm IG}\sim 10^{-20}{\rm G}$).}. As
an illustration, they have calculated the well-studied blazar
Mkn501, and obtained strong GeV emission, as long as the
intergalactic magnetic field (IGMF) is weak enough.

In this paper, we turn to investigate another important
extragalactic object, H1426+428, which has been detected by
Remillard et al. (1989) firstly. With the relatively high redshift
$z=0.129$, the optical depth exceeds unity for energies of
$\sim300$GeV (e.g. de Jager \& Stecker 2002). Any detection of a
signal at TeV energies translates directly into a high luminosity
of the source. The detection and spectral measurement of H1426+428
by the CAT, and VERITAS groups (Djanati-Atai et al. 2002; Petry et
al. 2002) indicate a steep spectrum between 250GeV and 1TeV,
whereas at higher energies, detections have been reported by the
HEGRA group based upon observations carried out in 1999 and 2000
(Aharonian et al. 2002b). Here, following Dai et al. (2002), we
study such hitherto undiscovered GeV emission by taking the
1998-2000 CAT data (Djannati-Ata\"{\i} et al. 2002), the
reanalyzed 1999 \& 2000 HEGRA data as well as the derived
intrinsic spectra proposed by Aharonian et al. (2003a).

The predicted GeV emission is sensitive to IGMF (Dai et al. 2002).
If the IGMF is stronger than $10^{-12}{\rm G}$, the formation of a
very extended electron/positron Halo ($R\sim10{\rm Mpc}$ for the
primary photons with a typical energy $\sim 1{\rm TeV}$) is
unavoidable (Aharonian, Coppi \& V\"{o}lk 1994), and its emission
is nearly isotropic. If this is the case, the GeV emission
predicted below will be significantly lowered. However, the
strength of IGMF has not been determined so far. Faraday rotation
measures imply an upper limit of $\sim 10^{-9}$ G for a field with
1 Mpc correlation length (see Kronberg 1994 for a review). In
dynamo theories, to interpret the observed $\mu$G magnetic fields
in galaxies and X-ray clusters, the seed fields required may be as
low as $10^{-20}$ G (see Kulsrud 1999 for a review). Theoretical
calculations of primordial magnetic fields generated during the
cosmological QCD or electroweak phase transition show that these
fields could be of order $10^{-20}$ G or even as low as $10^{-29}$
G (Sigl, Olinto \& Jedamzik 1997). The model proposed by
Aharonian, Timokhin \& Plyasheshnikov (2002) to explain the
observed TeV emission of Mkn501 suggests that $B_{\rm IG}\leq
10^{-18}{\rm G}$.  Modelling the long delayed MeV-GeV emission of
GRB 940217 suggests that $B_{\rm IG}\sim 10^{-20}{\rm G}$ (see
footnote 1). It is evident that all of these claims are far from
conclusive. Therefore, observing a hitherto undiscovered GeV
emission component from flares of TeV blazars such as H1426+428,
one may be able to obtain important information or constraints on
the poorly known IGMFs.

This paper is structured as follows: In section 2, we describe our
physical model briefly. In section 3 we provide our numerical
results. In section 4 we give our conclusions.

\section{Physical model}

As mentioned above, as the $\gamma-$rays with the energy high up
to $\sim 1{\rm TeV}$ travel toward the observer,  a significant
fraction of them is likely to be absorbed, yielding $e^{\pm}$. The
pair production optical depth $\tau_{\gamma\gamma}^{\rm ex}$
depends strongly on the $\gamma$-ray energy \footnote{In the
previous works (e.g. Dai et al. 2002), the red-shift correction of
$\gamma_{\rm e}$ as well as the four timescales involved (see
below) has been ignored. However, in the present case, for the
relatively high red-shift of H1426+428, $z=0.129$, the red-shift
correction must be taken into account.}
($E_\gamma'\equiv(1+z)E_{\rm \gamma,obs}$) and the redshift ($z$)
(throughout this letter, the subscripts ``obs" and ``loc" denote
the parameters measured in the observer and source frames
respectively). If a primary photon with energy $E_\gamma'$ has
been absorbed, the resulting $e^\pm$ pairs have Lorentz factors
$\gamma_e \simeq E_\gamma'/(2m_ec^2)\approx 10^6E_\gamma'/1\,{\rm
TeV})$, where $m_e$ is the electron mass. Such ultra-relativistic
$e^{\pm}$ pairs will subsequently Compton scatter on the ambient
CMBPs, and boost them to an average value $\sim
\gamma_e^2{\bar\epsilon}\simeq 0.63(1+z)(E_\gamma'/1\,{\rm
TeV})^2$ GeV, where ${\bar\epsilon}=2.7kT$ is the mean energy of
the CMB photons with $T\simeq 2.73(1+z)\,$K and $k$ is the
Boltzmann constant.

Since the emission of blazars is jet-like rather than isotropic,
so do the scattered CMBPs. Thus it is necessary to investigate
whether significant part of them has been scattered out of the
line of sight (the ``escaping'' effect), or not. For the IC-
scattered CMBPs, which move nearly along the path of the
ultra-relativistic electrons and the deflection angle is $\sim
1/\gamma_{\rm e}$, much less than the typical open angles of
blazars. On the other hand, the recoiling of the seed electrons is
small, too. It is easy to show that for one scattering, the
deflection angle is $<\bar{\epsilon}/m_{\rm e}c^2$. For a seed
electron scattering N times, its final deflection angle can be
estimated by $\theta_{D}\sim \sqrt{N} \bar{\epsilon}/m_{\rm
e}c^2$. Typically, for the IC scattering considered in this paper,
$N\approx1000$, which implies $\theta_{\rm D}\sim 10^{-7}$.
Therefore, the ``escaping'' effect can be ignored.

\subsection{The GeV emission duration}
As shown in Dai et al. (2002), there are four timescales involved
in the emission process: The first is $t_{\rm var,obs}$, the
observed variability time of the source emission, which has been
mentioned earlier.

The second is the well-known angular spreading time
\begin{equation}
\Delta t_{\rm A,obs}\approx(1+z){R_{\rm pair}\over 2\gamma_{\rm
e}^2c}=960(1+z)({\gamma_{\rm e}\over10^6})^{-2}({n_{\rm IR}\over
0.1{\rm cm^{-3}}})^{-1}{\rm s},
\end{equation}
where $R_{\rm pair}=(0.26\sigma_{\rm T}n_{\rm
IR})^{-1}\approx5.8\times 10^{25}({n_{\rm IR}\over 0.1{\rm
cm^{-3}}})^{-1}{\rm cm }$ is the typical pair-production distance,
$\sigma_T$ is the Thomson cross section, and $n_{\rm IR}\simeq
0.1{\rm cm^{-3}}$ is the intergalactic infrared photon number
density (Dai \& Lu 2002; Dai et al. 2002; Wang et al. 2003).

The third is the IC cooling timescale. In the absence of any IGMF,
the IC cooling timescale in the observer frame would be (Dai et
al. 2002)
\begin{equation}
\Delta t_{\rm IC,obs}\simeq (1+z)t_{\rm IC,loc}/(2\gamma_e^2)
=38(1+z)^{-3}(\gamma_e/10^6)^{-3}\,{\rm s}.
\end{equation}
$t_{\rm IC,loc}$, the IC cooling time scale measued in the source
frame is estimated by $t_{\rm IC,loc} =
3m_ec/(4\gamma_e\sigma_Tu_{\rm cmb})= 7.7 \times
10^{13}(1+z)^{-4}(\gamma_e/10^6)^{-1}\,{\rm s}$, where $u_{\rm
cmb}=aT^4$ is the CMB energy density, and $a$ is the radiation
constant.

The fourth is the magnetic deflection time, which appears in the
presence of IGMFs. The additional emission time due to magnetic
deflection is (Plaga 1995; Dai et al. 2002)
\begin{eqnarray}
\Delta t_{\rm B,obs} &\simeq & (1/2)(1+z)t_{\rm IC,loc}
\theta_{\rm B}^2\nonumber\\
&\simeq & 6.1\times 10^3({\gamma_e\over 10^6})^{-5}({B_{\rm
IG}\over 10^{-20}{\rm G}})^2(1+z)^{-11}\,{\rm s},
\end{eqnarray}
for $\theta_{\rm B}$, the deflection angle $\simeq \lambda_{\rm
IC}/R_{\rm L}=1.3\times 10^{-5}(\gamma_e/10^6)^{-2}(B_{\rm
IG}/10^{-20}{\rm G})(1+z)^{-4}\ll 1$, where $\lambda_{\rm
IC}\simeq c t_{\rm IC,loc}$, $R_{\rm L}=\gamma_{\rm e}m_{\rm
e}c^2/eB_{\rm IG}$ is the Larmor radius of the electrons, $B_{\rm
IG }$ is the strength of the IGMFs.

Consequently, the duration estimate of the IC emission from
electron/positron pairs scattering off the CMB is $\Delta t
(\gamma_e) = \max(\Delta t_{\rm IC,obs}, \Delta t_{\rm A,obs},
\Delta t_{\rm B,obs}, t_{\rm var,obs})$. Generally, the typical
duration of GeV emission is determined by the source activity time
and the energy dependent magnetic deflection time. For
convenience, we introduce a critical Lorentz factor $\gamma_{\rm
e,c}$ which is defined by $\Delta t_{\rm B,obs}= t_{\rm var,obs}$
\begin{equation}
{\gamma_{\rm e,c}\over 10^6}=0.59({t_{\rm var,obs}\over 1 {\rm
day}})^{-1/5}({B_{\rm IG}\over 10^{-20}{\rm
G}})^{2/5}(1+z)^{-11/5}.
\end{equation}
For electrons with Lorentz factor $\gamma_{\rm e}<\gamma_{\rm
e,c}$, the magnetic deflection angle is such large that most of
the scattered CMBPs can not reach the observer in a finite time
$t_{\rm var,obs}$, which contribute little to the GeV emission
considered here. Therefore, we take $\gamma_{\rm e,c}$ as the
lower limit for the integration performed below.

\subsection{Basic formulae for calculation}
For the ultra-relativistic seed electrons with the distribution
$dN_{\rm e}/d\gamma_{\rm e}$, the time-averaged scattered photon
spectrum is given by (Blumenthal \& Gould 1970; also see Dai \& Lu
2002; Dai et al. 2002)
\begin{equation}
\frac{dN_\gamma^{\rm SC}}{dE_{\gamma'}}={1\over 4\pi D_{\rm
L}^2}\int\int \left(\frac{dN_e}{d\gamma_e}\right)
\left(\frac{dN_{\gamma_e,\epsilon}}{dtdE_{\gamma'}}\right)dt
d\gamma_{\rm e},
\end{equation}
where $D_{\rm L}$ is the luminosity distance to the source,
$E_{\gamma'}$ is the externally scattered photon energy, $t$ is
the time measured in the local rest frame, and (Blumenthal \&
Gould 1970)
\begin{eqnarray}
{dN_{\gamma_e,\epsilon}\over dtdE_{\gamma'}}&=&{\pi r_0^2c\over
2\gamma_{\rm e}^4}{n(\epsilon)d\epsilon \over \epsilon^2}[2E_{\rm
\gamma'}{\rm ln}({E_{\gamma'}\over 4\gamma_{\rm e}^2
\epsilon})\nonumber\\
&& +E_{\gamma'}+4\gamma_{\rm e}^2\epsilon
-E_{\gamma'}^2/2\gamma_{\rm e}^2\epsilon],
\end{eqnarray}
is the spectrum of photons scattered (using the Tompson
cross-section formula) by an electron with Lorentz factor of
$\gamma_e$ from a segment of the CMBP gas of differential number
density (Blumenthal \& Gould 1970)
\begin{equation}
n(\epsilon)=[\pi^2(\hbar
c)^3]^{-1}[\epsilon^2/(e^{\epsilon/kT}-1)].
\end{equation}
The integration for variable $\gamma_{\rm e}$ ranges from
$\gamma_{\rm e,c}$ to $E'_{\rm up,cut}/2m_{\rm e}c^2$, where
$E'_{\rm up,cut}$ is the observed upper cut-off energy timed by
(1+z). Please note that this is only valid at the beginning since
$\gamma_{\rm e}$ decrease with $t$, which can be written into
$\gamma_{\rm e(t)}=\gamma_{\rm e,0}/(1+b\gamma_{\rm e,0}t)$, where
$\gamma_{\rm e,0}$ is the original Lorentz factor of the seed
$e^\pm$ pairs, $b=4\sigma_{\rm T} U_{\rm cmb}/3m_{\rm e}c$.

Dai \& Lu (2002) simply replace $\int dt$ by $t_{\rm IC,loc}$, and
assume the seed electrons have a distribution $\propto \gamma_{\rm
e}^{\rm -\Gamma_1}$. With these simplifications, they have shown
that in the Thomson limit, the scattered photon spectrum $\propto
E_{\rm \gamma'}^{-(\Gamma_1+2)/2}$ for $\Delta t_{\rm
B,obs}(\gamma_{\rm e })\ll t_{\rm var,obs}$ and $\propto E_{\rm
\gamma'}^{-(\Gamma_1-3)/2}$ for $\Delta t_{\rm B,obs}(\gamma_{\rm
e })\gg t_{\rm var,obs}$. Instead of taking such a simple
estimate, below we make a detailed numerical calculation: the
integration for variable $t$ ranges from 0 to $7.7\times
10^{13}(1+z)^{-4}(\gamma_{\rm e,c}/10^6)^{-1}{\rm s}$ .

\section{Numerical result}
In our calculation, the original seed electron energy spectrum is
derived as follows: we simply apply a polynomial fit to the local
highly absorbed photon spectrum: Setting
$x\equiv\log(E_\gamma'/1.0{\rm TeV})$, we have
\begin{equation}
{dN_{\rm \gamma}\over dE_\gamma'}\approx
10^{0.79046-3.31624x+1.15145x^2-0.79715x^3}
\end{equation}
which is in units of $4\pi D_{\rm L}^2 10^{-12}{\rm
ph~cm^{-2}~s^{-1}~TeV^{-1}}$. The local intrinsic photon spectra
$dN_{\rm \gamma, int}^{\rm i}/dE_{\rm \gamma'}$ (also in units of
$4\pi D_{\rm L}^2 10^{-12}{\rm ph~cm^{-2}~s^{-1}~TeV^{-1}}$) can
be approximated as follows (Aharonian et al. 2003a):
\begin{equation}
{dN_{\rm \gamma, int}^{\rm a}\over dE_\gamma'}\approx 5.8\times
10^3({E_\gamma'\over 0.1{\rm TeV}})^{-3/2},
\end{equation}
\begin{equation}
{dN_{\rm \gamma, int}^{\rm b}\over dE_\gamma'}\approx \left\{
\begin{array}{ll}
   1.6\times10^2({E_\gamma'\over {0.10\rm TeV}}),& {\rm E_\gamma'\leq 0.85TeV},\;\;\\
  1.4\times10^3({E_\gamma' \over {0.85\rm TeV}})^{-2},&  {\rm E_\gamma'\geq 0.85TeV}.\;\;\\
             \end{array} \right.
              \;\end{equation}
 \begin{equation}
{dN_{\rm \gamma, int}^{\rm c}\over dE_\gamma'}\approx \left\{
\begin{array}{ll}
   6.5\times 10^3({E_\gamma'\over {0.10\rm TeV}})^{-2},& {\rm E_\gamma'\leq 1.6TeV},\;\;\\
  25.4({E_\gamma'\over 1.60{\rm TeV}}),&  {\rm E_\gamma'\geq 1.6TeV}.\;\;\\
             \end{array} \right.
              \;\end{equation}
where the superscript $i=a,b,c$ represent the correction models
proposed by Primack et al. (2001), Aharonian et al. (2003a) and
Malkan \& Stecker (2001) respectively.  Since $E_\gamma'\approx
1(\gamma_{\em e}/10^6)~{\rm TeV}$, we have (including positrons)
\begin{equation}
{dN_{\rm e}^{\rm i}\over d\gamma_{\rm e}}\approx 2\times
10^{-6}({dN_{\rm \gamma, int}^{\rm i}\over dE_\gamma'}-{dN_{\rm
\gamma}\over dE_\gamma'})
\end{equation}
Now we can calculate the spectra of the scattered CMBPs with
equations (5)-(12). Please note that the $\epsilon$, $E_{\gamma'}$
in equations (6), (7) are all measured in the local frame rather
than by the observer.

\begin{figure}
\epsfig{file=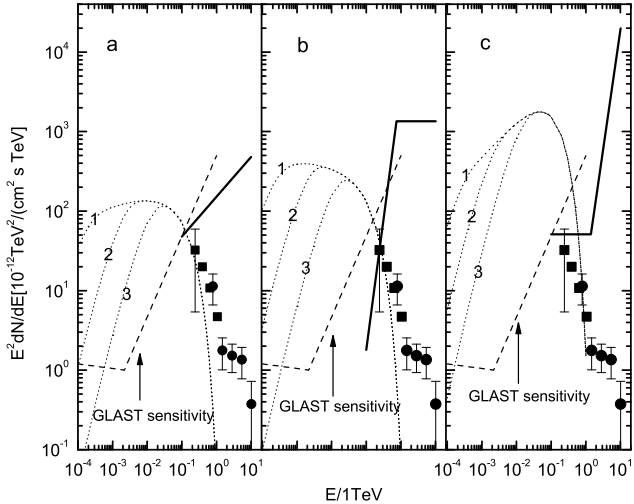, width=100mm} \caption{The time-averaged
high energy Gamma-ray spectra of H1426+428 for a flare with
$t_{\rm var,obs}=2~{\rm days}$. The filled circles are the
reanalyzed 1999/2000 spectral data measured by HEGRA (Aharonian et
al. 2003a), and the filled rectangles are the 1998-2000 spectral
data measured by the CAT (Djannati-Ata\"{\i} et al. 2002). In a, b
and c: the thick solid lines represent the derived intrinsic
spectra based on the correction model proposed by Primack et al.
(2001), Aharonian et al. (2003a) and Malkan \& Stecker (2001) (see
also figure 4 of Aharonian et al 2003a) respectively; the thick
dashed lines represent the GLAST sensitivity computed for an
exposure time of 2 days; the dotted lines are the secondary photon
spectra calculated in the text from the resulting $e^\pm$ pairs
interacting with the CMBPs, the number $1-3$ correspond to IGMF
strength of $10^{-20}{\rm G}$, $10^{-19} {\rm G}$ and
$10^{-18}{\rm G }$, respectively.}
\end{figure}

The numerical result has been shown in Figure 1. As long as the
IGMF strength is weak enough, there is the very strong GeV
emission (although it is intrinsic spectrum-dependent), whose flux
is far above the thick dashed line, the GLAST sensitivity computed
for an exposure time of 2 days. Just as shown in figure 1, the
inferred intrinsic spectra is most sensitive to the
characteristics of the extragalactic background light. As a
consequence, different correction models lead to much different
intrinsic spectra, so does the accompanying GeV emission, and thus
the detection of such emission may in turn provide us a valuable
chance to test those IR background absorption models.

In our calculations, as an illustration, we take $E'_{\rm
up,cut}=10.2~(1+z){\rm TeV}$ (Above which there is no observation
available), which is conservative. Even so, with the three
correction models taken in Aharonian et al. (2003a), the predicted
time-averaged scattered photon energy flux are far above (case b,
c), or at least comparable with (case a) that of the well studied
Blazar Mkn501 (see Dai et al. 2002 for detail). One problem
arises, just as shown in figure 1 c, the predicted new hard
$\gamma-$ray emission at the energy $\sim0.24~{\rm TeV}$ is about
ten times that observed. In principle, such emission may be
absorbed by the background IR/UV photons again, but the resulting
flux is still high above the observed (for $E_{\rm \gamma}\sim
0.24{\rm TeV}$, $\tau_{\gamma\gamma}^{\rm ex}\sim 1$). Two
possibilities exist: one is that the derived intrinsic spectrum
based on the correction model proposed by Malken \& Stecker (2001)
is much harder than what it really is. Another is that the IGMF
might be strong enough, the observable GeV-TeV emission flux has
been suppressed significantly.

For Blazar H1426+428, due to the poorly known X-ray band spectrum,
we cannot provide a definite SSC spectrum as Dai et al. (2002)
have done for Mkn501. Even so, as shown in Figure 1, at a first
glance, the SSC component is likely far below than the externally
scattered photon flux.

\section{Conclusions}

The HBLs H1426+428 is distinguished by its relatively high
redshift as well as its strong TeV energy emission. In this
Letter, with the recently observed data as well as the derived
intrinsic spectrum, we have studied the possible accompanying GeV
emission, which is due to the IC scattering of CMBPs by the
$e^\pm$ pairs produced in interactions of high energy photons with
the cosmic infrared-UV background photons. If the IGMF is weak
enough, there is very strong hitherto undiscovered GeV-TeV
emission, whose flux is far above or at least comparable with that
of the well-studied HBLs Mkn501 (Dai et al. 2002). With the
upcoming satellite GLAST, planned for launch in 2005, it is easy
for us to detect such emission directly. We also noted that the
shape of the inferred source spectra is most sensitive to the
characteristics of the extragalactic background light. A different
correction model leads to a different intrinsic spectrum, so does
the accompanying GeV emission (see figure 1 for detail). Thus the
detection of such emission may in turn provide us a valuable
chance to test those IR background absorption models.

In Aharonian et al. (2003b), the detection of TeV $\gamma-$rays
from the BL Lac 1ES1959+650 has been reported.  1ES1959+650 is
located at a redshift of $z=0.047$, providing an intermediate
distance between the nearby blazars Mkn421 and Mkn501, and the
much more distant object H1426+428. Without doubt, for such a
source, the observed TeV emission has been absorbed significantly
by the infrared-UV background photons, and thus a strong GeV
emission is expected, as long as the IGMF is weak enough.

\acknowledgements We would like to thank Drs Y. F. Huang, X. Y.
Wang \& X. F. Wu for valuable comments. We also thank the
anonymous referee for her/his useful suggestions that enable us to
improve the paper significantly. This work was supported by the
National Natural Science Foundation of China (grants
10073022,10225314 and 10233010), the National 973 Project (NKBRSF
G19990754) and the Foundation for the Author of National Excellent
Doctoral Dissertation of P. R. China. (Project No:200125).

\end{document}